\documentclass[10pt,twocolumn]{article}

\usepackage{isqcmc2023}
\usepackage{graphicx,url}

\usepackage{cite}
\usepackage{amsmath,amssymb,amsfonts}
\usepackage{graphicx}
\usepackage{textcomp}
\usepackage{xcolor}
\usepackage{algorithm}
\usepackage{algpseudocode}
\usepackage{subfig}
\usepackage{multirow}
\usepackage{adjustbox}
\usepackage{hyperref}

\newcommand{\linebreakand}{%
  \end{@IEEEauthorhalign}
  \hfill\mbox{}\par
  \mbox{}\hfill\begin{@IEEEauthorhalign}
}
\makeatother 

\def\BibTeX{{\rm B\kern-.05em{\sc i\kern-.025em b}\kern-.08em
    T\kern-.1667em\lower.7ex\hbox{E}\kern-.125emX}}
\begin{document}

\sloppy

\title{Quantum Memory of Musical Compositions}

\author{Maria Mannone\inst{1}\inst{2} \and
        Omar Costa Hamido\inst{3}}

\address{Dipartimento di Ingegneria, Università di Palermo,  Italy
\nextinstitute
DSMN - ECLT, Università Ca' Foscari di Venezia, Italy
\nextinstitute
University of Coimbra, Centre for Interdisciplinary Studies (CEIS20), Coimbra, Portugal \email{maria.mannone@unive.it, ocostaha@uci.edu}\\
\url{http://mariamannone.com}, \url{https://omarcostahamido.com}
\\
}

\maketitle

\begin{abstract}
The 
\textcolor{black}{perception and appreciation} of 
beauty in a musical composition appears as being related to the amount of memory and balance between expectation and surprise. In this study, we use a formal computing tool derived from quantum mechanics and adapted to music, to measure the degree of memory \textcolor{black}{of orchestral music}, including \textit{Liebestod} by R. Wagner. We discuss our results and further lines of research.
\end{abstract}


\section{Introduction}

What is more pleasurable, a soft music, or an unexpected chord which creates tension, lately released into a consonance? In one of his talks at Harvard,\footnote{\protect\url{https://www.youtube.com/watch?v=A7O5zcQPRQQ}} Bernstein analyzed the beginning of the \textit{Adagietto} from Mahler's Fifth Symphony, with the initial (and thematic) tension of the dominant, quickly released, and the subsequent equilibrium of tension and relaxation throughout the movement---and the sense of ambiguity is related with the unique beauty of this piece. The affective reactions to harmonic, melodic, rhythmic, and orchestration tricks stimulated the curiosity of scientists. Some recent research focuses on the role of surprise and expectation across different listeners and large databases of musical compositions, including modern songs. The physiological responses of the brain are compared with the verbal replies of listeners, which are in turn compared with the characteristics of the musical pieces \cite{surprise, huron}. The perception \textcolor{black}{(or, more precisely, the appreciation)} of \textit{beauty} is probably related to some sort of balance between surprise and expectation, as in an indeterminacy relation \cite{gestART}.
\textcolor{black}{The problem with most works in this topic, with some noticeable exceptions such as \cite{surprise}, is the lack of hard experimental or empirical evidence to prove the beauty of whole artwork. Instead, we should look for the saliency of basic elements or essential principles - in music, they can be repetition, organization, symmetry, and surprise.}
Even though the concept of beauty and aesthetic pleasure can vary across listeners and cultures, we nevertheless believe that there is a \textit{universal core} of content that is shared, maybe related to the general way how the brain works, and our biological, natural roots--and we share them with non-human animals \cite{heinrich}. Maybe the seek for beauty is related to survival itself \cite{prum}. Maybe it is also related to the formal organization of an artwork. Here, we try to use computational tools to characterize the information content of a musical artwork, i.e. formal grammars to formalize thematic organization  \cite{blutner}.

\textcolor{black}{The topic of beauty in nature and in the arts is vast. Here, we focus on a smaller portion of this field, that is, the balance between expectation and surprise. In particular, we consider the \textit{degree of memory} as an essential element of understandability and appreciability of an artwork which unfolds through time. Memory is related with the sequence of events on time and their similarity. In a musical framework, this is related with patterns of pitches, rhythms, timbres, and loudness. The degree of memory in a musical piece can be computed.} 
\textcolor{black}{The amount of memory in a musical composition has been measured using a criterion developed in physics, and adapting it to music \cite{non_markov_music_paper}.}
\textcolor{black}{In this paper, we present a preliminary analysis of memory in musical pieces using}
a quantum-derived measure of the amount of memory. We compare our empirical judgments on similarity/change inside these pieces with the information that can be recovered from these techniques.
\textcolor{black}{The computational advantages introduced by quantum computing can be,} 
\textcolor{black}{and are more often} \textcolor{black}{applied to research in music information retrieval \cite{putz}; quantum simulators are used to create music \cite{rocchesso, miranda}; the dichotomy between continuous and discrete, one of the principles of quantum mechanics, can lead to deep investigation in all objects involved in music, from the continuum of pitches to the discreteness of scale tones and chords \cite{fugiel} and the modeling of music-theory concepts in terms of forces \cite{graben}. For all these reasons,}
quantum mechanics and quantum computing can provide new insights for music analysis. \textcolor{black}{Our study is a piece in this picture: how a quantum physics-based criterion can be transferred and adapted to the domain of classical physics, in particular musical applications, helps us investigate these criteria's limitations and advantages, as well as lead to objective measurements of parameters related to our perception.}

The article is organized as follows. In Section \ref{methods}, we present our formal tool of analysis, to quantify the level of novelty and expectation. In Section \ref{results}, we apply the tools to the chosen musical compositions, and we comment on the obtained results. In Section \ref{conclusions}, we summarize our research and provide some hints for its future developments.

\section{Method}\label{methods}


We consider here the degree of \textcolor{black}{musical} non-Markovianity \cite{non_markov_music_paper}.\footnote{\textcolor{black}{The application of non-Markovianity quantifiers to music was proposed by the theoretical physicist Giuseppe Compagno}.} It is an adaptation of a quantifier of similarity for quantum states \cite{non-markov_physics}, \textcolor{black}{in the framework of open quantum systems \cite{cover_thomas, preskill, breuer}.} In physics, if the distinguishability between two quantum states is kept over time, then the memory remains \cite{breuer} and the dynamics is non-Markovian \cite{plenio}. In music, it is the opposite: the lesser the changes between two consequent musical sequences, the higher the thematic memory. \textcolor{black}{In physics \cite{breuer}, the rate $\sigma$ between two states described by density matrices $\rho_1,\rho_2$ is defined as: \begin{equation}
\sigma = \sigma(\rho_1,\rho_2) = \frac{d}{dt}D(\rho_1,\rho_2)=\frac{d}{dt}\left(\frac{1}{2}tr|\rho_1-\rho_2|\right).
\end{equation}
In the musical case, the matrices $\rho_i$ are distribution matrices of pairs of parameters in finite musical sequences. In fact,
}
\textcolor{black}{while in physics the comparisons are computed between different states at the same instant, in music we compare sequences of finite duration.}

\textcolor{black}{First,} for each musical sequence, we compute distribution matrices \textcolor{black}{concerning pairs of parameters}. Then, for corresponding matrices (e.g., the \texttt{frequency-start} matrices of two sequences), we compute their differences, obtaining the rates $\sigma_i$. 
The non-Markovianity degree for, let's say, \texttt{frequency-start} matrices, is defined as:
\begin{equation}\label{degree}
\mathcal{M}_C=\frac{1}{1+r\sum_i(if\,\sigma_i>0)\sigma_i},
\end{equation}
where $r=n^+/n^T$ is the fraction of positive rates over the total rates. See \cite{non_markov_music_paper} for the proof of Eq. \ref{degree}.

\begin{figure}[ht!]
\centerline{\includegraphics[width=0.5\columnwidth]{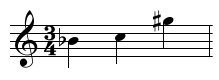}}
\caption{\textcolor{black}{A simple musical sequence, from which the matrix of eq. \ref{matrix1} is derived.}}
\label{seq1}
\end{figure}

\textcolor{black}{We show here two examples of music-derived density matrices. The first one was proposed in \cite{non_markov_music_paper}. The musical fragment of Figure \ref{seq1} leads to the start-frequency matrix of Eq. \ref{matrix1}. To obtain this matrix,\footnote{See also the section on \textbf{Availability of codes}.} we first need to work on pitches and on onsets separately. In short, we evaluate the normalized distance between each note, and the mean value (of pitch, and then of onsets) of the sequence. Then, we count the number of notes belonging to a given interval of pitch (and then of onset), in the order
$[0, 0.25]$, $[0.25, 0.5]$, $[0.5, 0.75]$, $[0.75, 1]$. Finally, we build the matrix having the interval of onsets on the rows, and of pitches on the columns, and we normalize the whole matrix, dividing each cell by the total number of notes in the sequence, in order to have $1$ as the sum of all the elements of the sequence.}

\begin{equation}\label{matrix1}
\rho_1=
\left(
\begin{matrix}
0.33 & 0 & 0 & 0 \\
0 & 0 & 0 & 0 \\
0 & 0 & 0 & 0.33 \\
0 & 0 & 0 & 0.33 \\
\end{matrix}
\right)
\end{equation}

\begin{figure}[ht!]
\centerline{\includegraphics[width=\columnwidth]{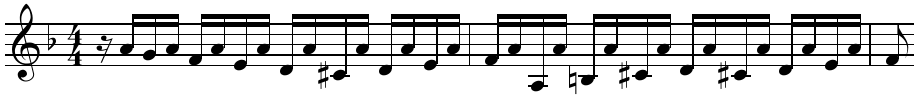}}
\caption{\textcolor{black}{J. S. Bach, \textit{Toccata and Fugue in D minor}, BWV 565, beginning of the fugue. The corresponding \texttt{start-intensity} matrix is presented in Eq. \ref{matrix2}.}}
\label{bach_fugue}
\end{figure}

\textcolor{black}{The second example is the opening sequence of the Fugue from \textit{Toccata and Fugue in D minor}, BWV 565, by J. S. Bach (Figure \ref{bach_fugue}). The corresponding start-frequency matrix is shown in Eq. \ref{matrix2}.}

\begin{equation}\label{matrix2}
\rho_2=
\left(
\begin{matrix}
0.06 & 0 & 0.03 & 0.12 \\
0.12 & 0.19 & 0.19 & 0.12 \\
0 & 0.06 & 0.03 & 0.00 \\
0.06 & 0 & 0 & 0 \\
\end{matrix}
\right)
\end{equation}




\section{Results}\label{results}

In \cite{non_markov_music_paper}, three pieces were compared, with very different degrees of memory: \textit{Dolente immagine} by V. Bellini, \textit{Solo} for oboe by B. Maderna, and \textit{Metamorphosis III} by P. Glass.

Here, we compare the results which can be obtained for two longer orchestral musical pieces.
The first one is the well-known \textit{Liebestod} from \textit{Tristan und Isolde} by R. Wagner. The second is a contemporary orchestral piece, dedicated to St. Rosalia and her relation with the city of Palermo by M. Mannone, whose premiere recording can be accessed online.\footnote{\protect\url{https://soundcloud.com/maria-mannone/santa-rosalia}} 
The first piece is characterized by a unitary structure, with progressive variations, and the typical \textit{infinite melody}\footnote{\textcolor{black}{The \textit{endless melody} in Wagner pieces refers to the concatenation of melodic variation of some initial material, constituting a continuous flow, that lacks separation between melodic sequences.}} by Wagner. The second piece is in the form of program music, with shorter, descriptive sequences and short \textit{leitmotiv} that are recalled in several points.

While in \cite{non_markov_music_paper}, for the matrix computation, the pieces were divided into short sequences, here we halved them. The two sets of matrices corresponding to the two halves are then used to compute the degree of non-Markovianity. We chose this strategy as an experiment to get an overall picture of the pieces. \textcolor{black}{This is also the first time the musical non-Markovianity degrees are computed for orchestral scores.} In our study, the coefficients provide a measure of the stylistic unit between the first and the second part of each composition.
\textcolor{black}{We also consider the values obtained with an improvisation,
\textcolor{black}{In the following, we will be using the term ``memory degree'' as a synonym of ``non-Markovian degree.''}

\begin{table}[ht!]
\centering
\begin{tabular}{cccc}
\hline
$M_c$ coefficient & Tristan & Rosalia & improvisation \\
\hline
frequency - start & 0.9390 & 0.9132 & 0.8850 \\
frequency - duration & 0.9950 &  0.9524 & 0.8734 \\
frequency - intensity & 0.8475 & 0.9479 & 0.9390 \\
duration - intensity & 0.7722 & 0.9950 & 0.8264 \\
start - intensity & 0.9259 & 0.9091 & 0.8850 \\
\hline
\end{tabular}
\caption{Non-Markovianity coefficients for each typology of musical matrices, computed for the \textit{Tristan},  \textit{Santa Rosalia}, and a two-part improvisation.}\label{non_M_table}
\end{table}


Observing the results reported in Table \ref{non_M_table}, we notice that the values of \textcolor{black}{thematic memory are higher for \textit{Tristan}, as expected, given the \textit{infinite melody} with the progressive transformation of the material. However, the \texttt{duration-intensity} memory degree is lower for \textit{Tristan}: This information takes into account the loudness change and climax in the second half of the piece, corresponding to a high expressive moment.
The variations in intensity are more homogeneously distributed in \textit{Rosalia}, and thus the \texttt{duration-intensity} memory degree shows a higher value. Concerning the improvisation, rhythm, and pitch distribution of the improvisation differs between the first and the second part; thus, the \texttt{frequency-start} memory value is lower than the corresponding degree for \textit{Tristan} and \textit{Rosalia}. The same goes with \texttt{start-intensity} memory degree, as the first part of the improvisation presented ostinato patterns and overall a different organization concerning the second one. Thus, the degree of memory is different. However, there is a key difference between the improvisation \textcolor{black}{short takes} and the other considered pieces: the orchestral pieces present a more variegated pitch distribution over time, with multiple lines, and superposed rhythms; these elements contribute to the relative lowering of the memory degree concerning the improvisation.}
\textcolor{black}{For a numerical comparison, we can consider that, in \cite{non_markov_music_paper}, a value of $0.3$ for memory degree was obtained comparing a measure with a quasi-random melodic and a completely different measure, constituted only by a whole note.}

\begin{figure}[ht!]
\centerline{\includegraphics[width=\columnwidth]{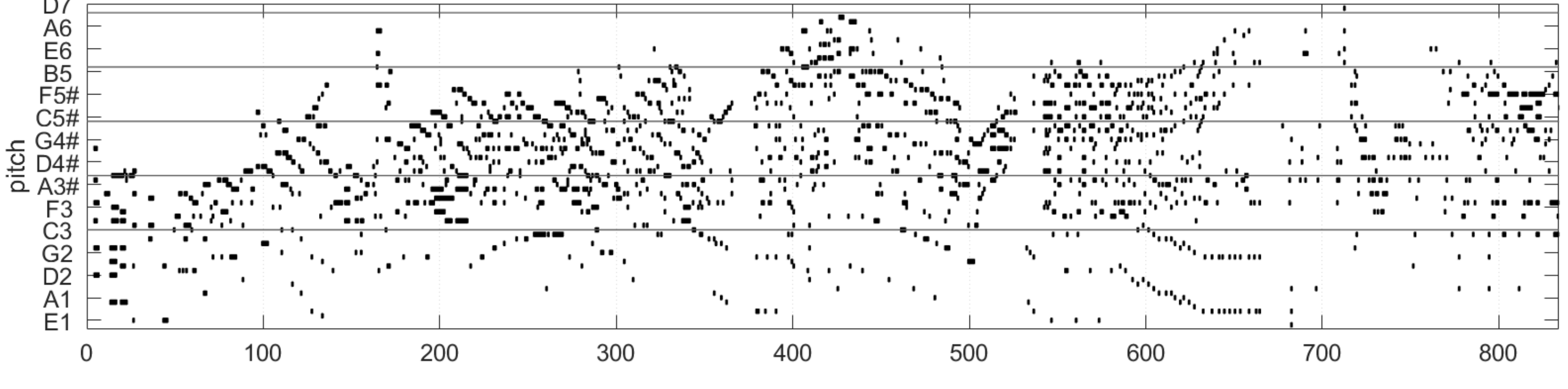}}
\caption{\textit{Liebestod} from \textit{Tristan und Isolde}, piano-roll notation.}
\label{tristan}
\end{figure}

\begin{figure}[ht!]
\centerline{\includegraphics[width=\columnwidth]{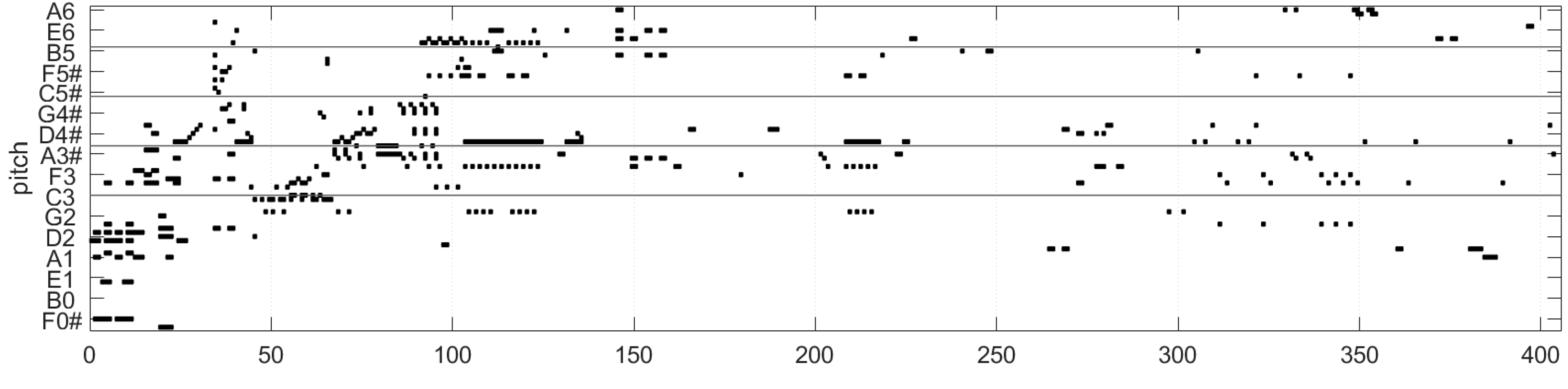}}
\caption{\textit{Santa Rosalia}, piano-roll notation.}
\label{rosalia}
\end{figure}

\section{Discussion and Conclusions}\label{conclusions}

In this article, we considered a quantum-based quantifier, and applied it to the measurement of thematic memory in orchestral pieces.

The next step of the research \textcolor{black}{will include feedback from listeners. In fact, we can explore the level of concordance between computed memory degrees and qualitative judgments of a large sample of listeners. We could also try to understand if} the quantifiers can highlight some information non-retrievable \textcolor{black}{from a professional or non-professional listening.}
\textcolor{black}{Proper experimental validation of our claims could allow us to deepen the interpretation of the results we obtained, and of the future ones to be obtained.}

\textcolor{black}{Further research can investigate the degrees of memory of smaller sections of the pieces, comparing the results we can obtain locally with some salient points present in the score, and some elements noticed by listeners. In this way, we will able to assess with more precision the effect of memory degree on the perception of musical interest, and on the eliciting of attention and aesthetic pleasure.}
\textcolor{black}{Global and local non-Markovianity coefficients can help assess the approximated interest, at least concerning the degree of \textit{surprise} and \textit{expectation}. The proposed strategy may even become helpful to deal with the virtually infinite pool of generative music!
This could offer quantum composers an opportunity to evaluate the interest of the output musical material without the need for long hours of listening.}

Maybe science can help understand why something can be perceived as \textit{beautiful}, and maybe, with the right mapping, and understanding, it can also help create new \textit{beauty} in the world.



\section*{Availability of codes}

The original code for musical-matrix computation was written by M. Mannone in C, for \textcolor{black}{a chapter of her MSc in theoretical physics} (2012). It was then used for the article \cite{non_markov_music_paper}, \textcolor{black}{available as a preprint since 2013, and published in 2022.} The code was later ported to Max/MSP by Omar Costa Hamido in 2023 \cite{gitrepo},
making it more interactive, compatible with regular MIDI files as well as live input, and optimizing the workflow.

After having obtained the musical matrices, we computed the non-Markovianity degrees with MatLab. The technique was the following:
\texttt{[letter] = readmatrix(``path of the .txt file'')}, for matrices \texttt{frequency\_start}, \texttt{frequency\_duration}, \texttt{frequency\_intensity}, \texttt{duration\_intensity}, and \texttt{start\_intensity} of each of the parts a musical piece is divided into. We obtain matrices \textcolor{black}{A-B-C-D-E-F-G-H-I-L, following the Italian alphabet.}
Then, we consider pairs of corresponding matrices, e.g., the \texttt{frequency\_start} matrices $A$ and $B$, for the first and the second part of the piece, respectively. We compute: $AB = abs(A-B)$, $DAB = 0.5*trace(AB)$.
The corresponding non-Markovianity degree was then computed as: $1/(1 + DAB)$.



\begin{thebibliography}{00}



\bibitem{surprise}
V. K.M. Cheung, P. M.C. Harrison, L. Meyer, M. T. Pearce, J.-D. Haynes, S. Koelsch.
Uncertainty and Surprise Jointly Predict Musical Pleasure and Amygdala, Hippocampus, and Auditory Cortex Activity. \textit{Current Biology}, 29 (23): 4084--4092.e4, 2019

\bibitem{huron}
D. Huron. Musical Aesthetics: Uncertainty and Surprise Enhance Our Enjoyment of Music. Current Biology Dispatches, 29, R1224--R1251, 2019

\bibitem{gestART}
M. Mannone, F. Favali, B. Di Donato, L. Turchet. Quantum GestART: identifying and applying correlations between mathematics, art, and perceptual organization. Journal of Mathematics and Music, 15 (1): 62--94, 2020

\bibitem{heinrich}
B. Heinrich, The Biological Roots of Aesthetics and Art. \textit{Evolutionary Psychology}, 11(3), 2013 

\bibitem{prum}
R. O. Prum. \textit{The Evolution of Beauty: How Darwin's Forgotten Theory of Mate Choice Shapes the Animal World -- and Us}. 2017

\bibitem{blutner}
R. Blutner, \textit{Pragmatics and Lexicon}. \protect\url{http://www.blutner.de/pragmatics.pdf} 

\bibitem{graben}
P. beim Graben and R. Blutner. Gauge models of musical forces. \textit{Journal of Mathematics
and Music}, 15 (1): 17–36, 2020 

\bibitem{fugiel}
B. Fugiel. Quantum-like melody perception. \textit{Journal of Mathematics and Music}, 17(2): 319-331, 2022

\bibitem{miranda}
E. R. Miranda Quantum computer: Hello, music! In E. R. Miranda (Ed.), \textit{Handbook of artificial intelligence for music: Foundations, advanced approaches, and developments for creativity}. Springer, 2021

\bibitem{putz}
V. Putz and K. Svozil. Quantum music. \textit{Soft Computing}, 21, 1467–1471, 2017

\bibitem{rocchesso}
D. Rocchesso and M. Mannone. A quantum vocal theory of sound. \textit{Quantum Information
Processing}, 19 (292): 1–28, 2020

\bibitem{non_markov_music_paper}
M. Mannone and G. Compagno. Characterisation of the Degree of Musical Non-Markovianity. \textit{Journal of Creative Musical Systems}, 6(1), 2022

\bibitem{non-markov_physics}
Mannone, M., Lo Franco, R., Compagno, G. (2013). Comparison of non-Markovianity criteria in a
qubit system under random external fields. \textit{Physica Scripta T}, 153, 014047.

\bibitem{breuer}
H. P. Breuer and F. Petruccione. \textit{Theory of open quantum systems}. Oxford University Press, 2002

\bibitem{cover_thomas}
T. M. Cover and J. A. Thomas, J. A. \textit{Elements of information theory}. Wiley, 2006

\bibitem{preskill}
J. Preskill. \textit{Lecture notes for physics 219: Quantum computation}. California Institute of Technology, 2004

\bibitem{plenio}
\'{A}, Rivas, S. F. Huelga, M. B. Plenio. Quantum non-Markovianity: characterization, quantification and detection. \textit{Reports on Progress in Physics} 77 (9): 094001, 2014

\bibitem{gitrepo}
O. C. Hamido and M. Mannone, \textit{musical\_non-Markovianity} (v1.0.0). Zenodo. \protect\url{https://doi.org/10.5281/zenodo.8152691}, 2023



\end{thebibliography}

\end{document}